\documentclass[aps,prb]{revtex4-1}
\usepackage{bm}
\usepackage{amsmath}
\usepackage{amssymb}
\usepackage{amsfonts}
\usepackage{graphicx}
\usepackage[pdffitwindow,colorlinks,citecolor={red},linkcolor={blue}]{hyperref}

\begin{document}
\title{Second-Harmonic Generation in Nano-Structured Metamaterials}
\author{Ulises R. Meza}
\author{Bernardo S. Mendoza}
\affiliation{Department of Photonics, Centro de Investigaciones en \'Optica,
 Le\'on, Guanajuato, M\'exico}
\author{W. Luis Moch\'an}
\affiliation{Instituto de Ciencias F{\'i}sicas, Universidad Nacional 
 Aut\'onoma de M\'exico, Apartado Postal 48-3, 62251 Cuernavaca,
 Morelos, M\'exico.}

\begin{abstract}
We conduct a theoretical and numerical study on the second-harmonic
(SH) optical 
response of a nano-structured metamaterial composed of a periodic array of
inclusions. Both the inclusions and their surrounding matrix are made of
centrosymmetrical materials, for which SH is strongly suppressed, but by
appropriately choosing the shape of the inclusions, we may produce a
geometrically non-centrosymmetric system which does allow efficient SH
generation. 
Variations in the geometrical configuration allows tuning the
linear and quadratic spectra of the optical response of the
system. We develop an efficient scheme for calculating the nonlinear
polarization, extending a formalism for the calculation 
of the macroscopic dielectric function using Haydock's recursion
method. We apply the formalism developed here to an array of holes within an Ag
matrix, but it can be readily applied to any metamaterial made of arbitrary
materials and for inclusions of any geometry within the long-wavelength
regime.
\end{abstract}

%\pacs{42.65.-k, 78.67.Pt, 81.05.Xj, 78.20.Bh, 42.81.Gs}
%keywords: nonlinear optics, second harmonic generation,
%nanoparticles, metamaterials, epioptics, plasmonics 
\maketitle 

%%%%%%%%%%%%%%%%%%%%%%%%%%%%%%%%%%%%%%%%%%%%%%%%%%%%%%%%%%%%%%%%%%%%%%%%%%%%%%%% 
%%%%%%%%%%%%%%%%%%%%%%%%%%%%%%%%%%%%%%%%%%%%%%%%%%%%%%%%%%%%%%%%%%%%%%%%%%%%%%%%

\section{Introduction}

The advent of structured metamaterials has allowed the design of new
materials, with an unprecedented amount of control over their intrinsic
properties. These metamaterials are typically composite systems that consist of
two or more ordinary materials, that are periodically structured or arranged in
such a manner that the resulting properties differ from those of the constituent
materials. These systems have been widely explored both theoretically and
experimentally, with a plethora of new applications under development
\cite{veselagoUPS67, garlandAIP78, smithPRL00, smithSCIENCE04,husuNL12,
laroucheOC10}. The variety of available fabrication techniques such as
electron-beam lithography \cite{akahaneNATURE03, grigorenkoNATURE05,
balciJSTQE17}, ion milling \cite{gordonPRL04, seniutinasAPA16}, and even
conventional 3D printing \cite{wegenerLAMOM08, shenJO17, mikheevaAOTF18}, allow
for extremely precise designs of structured systems featuring arrays of
inclusions (or holes) with specific shapes. These methods allow the
fabrication of new devices with highly tunable optoelectronic properties
\cite{pendryPRL00, smithPRL00}. A wide variety of applications using
metamaterials have now been developed. Materials can be designed to have a
negative index of refraction \cite{shalaevOL05}; this has been implemented using
periodic noble metal inclusions within a dielectric matrix
\cite{kildishevJOSAB06}. Flat lens-like devices can be fabricated using
metamaterials that can manipulate the propagation of light with sub-wavelength
focusing capabilities; \cite{pendryPRL00} this type of device has been
implemented for cloaking \cite{pendrySCIENCE06, leonhardtSCIENCE06, haoAPL10}
and shielding applications \cite{fengPRL08}. The fabrication of these materials
is not restricted to specific ranges of the electromagnetic spectrum,
permitting, for example,  
the development of new devices designed to work in the terahertz regime
\cite{alekseyevAO12, bornEL15, suzukiOE18}.

Metamaterials display a wide variety of optical phenomena \cite{chenNM10}; of
particular interest to us are their nonlinear optical properties.
The nonlinear response is strongly sensitive to
the natural atomic structure; for second-harmonic generation (SHG), the material
must have a non-centrosymmetric crystalline structure in order to have a strong
dipolar nonlinear response. Structured metamaterials, that can be designed with
almost limitless configurations, make for a promising alternative for nonlinear
optical applications. There have been numerous theoretical \cite{laroucheOC10,
obrienNM15, larouchePRA18} and experimental \cite{shadrivovJOSAB06, fengPRL08,
husuNL12} studies concerning the development of nonlinear devices using
metamaterials. Some examples of nonlinear metamaterials have been fabricated
using split-ring resonators \cite{zharovPRL03, kleinSCIENCE06} and nano-rod
inclusions \cite{marinoLPR18}, producing SHG-active, magnetic, and left-handed
materials. Other inclusions can be intrinsically noncentrosymmetric
\cite{canfieldNL07}, thus creating a strong SHG response. Tailored metamaterials
allow for the possibility to tune the nonlinear optical response
\cite{chenNANOPHOTONICS12, timbrellSR18, barSI18, galantySA18} as a function of
the geometrical configuration. These systems can be varied geometrically,
changing their degree of non-centrosymmetry, thus allowing for the
second-harmonic (SH) signal to be enhanced.

The required physical parameters (namely, the electric permittivity
and magnetic 
permeability) that are used for calculating the linear optical
response can be obtained 
via a homogenization procedure \cite{smithPRL00, simovskiJO10,
aluPRB11}. The formalism presented in Refs. \onlinecite{mochanPRB85a} and
\onlinecite{mochanPRB85b} is used in this work to describe the macroscopic
linear response of inhomogeneous systems in terms of an average of
certain specific 
microscopic response functions of
the system. These quantities can then be used and the formalism may be
extended to calculate the linear and
non-linear optical responses of metamaterials of arbitrary composition
\cite{cortesPSSB10, mochanOE10, perezNJP13, mochanLAOPC14, mendozaPRB16}. In
this work, we explore the nonlinear SH response of a periodic nanostructured
metamaterial comprised of an array of holes of a non-centrosymmetric
geometry within a matrix made of a centrosymmeteric material, for
which we chose  silver. In this case, the SH generation from a
homogeneous matrix would be strongly suppressed, but the
noncentrosymmetric geometry of the holes allows a strong signal whose
resonances may be tuned and enhanced through variations of the geometrical
parameters\cite{butetJOSAB30, butetNANO9}. We systematically study the evolution 
of the nonlinear susceptibility tensor due to variations in the shape and
position of the holes. Lastly, we elucidate the
origin of the produced SH response by calculating and analyzing the charge
density and polarization field at the metallic surface.

The paper is organized as follows. In Sec. \ref{sec:theory} we present the
theoretical approach used to calculate the dielectric response of the
metamaterial that is then used to obtain the nonlinear SH polarization. In Sec.
\ref{sec:results} we present results for a nanostructured metamaterial
consisting of empty holes within a silver matrix. We explore a variety of
geometric configurations to fine-tune the SH response. Finally, in Sec.
\ref{sec:conc} we present our conclusions.

%%%%%%%%%%%%%%%%%%%%%%%%%%%%%%%%%%%%%%%%%%%%%%%%%%%%%%%%%%%%%%%%%%%%%%%%%%%%%%%%
%%%%%%%%%%%%%%%%%%%%%%%%%%%%%%%%%%%%%%%%%%%%%%%%%%%%%%%%%%%%%%%%%%%%%%%%%%%%%%%%

\section{Theory}\label{sec:theory}

The quadratic polarization forced at the second-harmonic (SH) frequency
$2\omega$ by an inhomogeneous fundamental field $\bm E_\omega$ at frequency
$\omega$ within an isotropic centrosymmetric material system made of
polarizable 
entities within the non-retarded regime may be written as \cite{jacksonbook}
\begin{equation}\label{Pf}
  \bm P^{f}({2\omega})=n\bm p({2\omega})-\frac{1}{2}\nabla\cdot n\bm
  Q({2\omega}) 
\end{equation}
where $n$ is the number density of polarizable entities,  $\bm p({2\omega})$ is
their electric dipole moment, given within the {\em dipolium}
model\cite{mendozaPRB96} by
\begin{equation}\label{p}
  \bm
  p({2\omega})=-\frac{n}{2e}\alpha(\omega)\alpha(2\omega)\nabla E^2(\omega),
\end{equation}
$\bm Q({2\omega})$ is their electric quadrupole moment, given by
\begin{equation}\label{Q}
  \bm Q({2\omega})=\frac{1}{2e}n\alpha^2(\omega)\bm E(\omega)\bm E(\omega),
\end{equation}
and $\alpha(\nu\omega)$ are the the linear polarizabilities of each entity at
the fundamental ($\nu=1$) and at the SH ($\nu=2$), related to the dielectric
function $\epsilon(\nu\omega)$ through
\begin{equation}\label{eps}
\epsilon(\nu\omega)=1+4\pi n \alpha(\nu\omega).
\end{equation}
We allow the density $n$, the polarizability $\alpha$, the dielectric response
$\epsilon$ and the field to depend on position. The total polarization induced
at the SH is then
\begin{align}\label{P}
  \bm P({2\omega})=&n\alpha(2\omega) \bm E({2\omega}) +\bm
  P^f({2\omega})\nonumber\\
  =&n\alpha(2\omega) \bm E({2\omega}) - \frac{n}{2e} 
  \alpha(\omega)\alpha(2\omega) \nabla E^2(\omega)
  +\frac{1}{2e}\nabla\cdot n\alpha^2(\omega)\bm E(\omega)\bm E(\omega),
\end{align}
where we added to Eq. (\ref{Pf}) the polarization linearly induced by the
self-consistent electric field $\bm E({2\omega})$ produced by the total
SH polarization $\bm P({2\omega})$.

We want to apply the equations above to obtain the nonlinear susceptibility of a
binary metamaterial consisting of a host made up some material $A$ in which
inclusions made up of a material $B$ are embeded forming a periodic
lattice. In our actual calculations we will replace material B by vacuum.
We
denote by $\epsilon_\gamma$, $\alpha_\gamma$ and $n_\gamma$ the dielectric
function, polarizability and number density corresponding to material
$\gamma=A,$ $B$. We may describe the geometry of the metamaterial through a
periodic {\em characteristic function} $B(\bm r)=B(\bm r+\bm R)$ which takes the
values 1 or 0, according to whether the position $\bm r$ lies within the region
occupied by material $B$ or $A$, respectively, and where $\bm R$ is a lattice
vector. Thus, we may write the dielectric function as
\begin{equation}\label{epsVsB}
  \epsilon(\bm r)=\frac{\epsilon_A}{u}(u-B(\bm r)),
\end{equation}
where we introduced the spectral variable
\begin{equation}\label{u}
  u=\frac{1}{1-\epsilon_B/\epsilon_A},  
\end{equation}
which takes complex values in general and accounts for the composition of the
materials and for their frequency dependent response.

In the long wavelength approximation, assuming that the unit cell of the
metamaterial is small compared to the wavelength of light in vacuum and the
wave- or decay-length within each of its components, we may take the electric
field within a single cell as longitudinal $\bm E=\bm E^L$ and we may identify
the longitudinal part $\bm D^L$ of the displacement field $\bm D$ as an {\em
external} field, which therefore has no fluctuations originated in the spatial
texture of the metamaterial, and is thus a macroscopic field $\bm D^L=\bm
D^L_{M}$. Thus, if we excite the system with a longitudinal external field we
may write
\begin{equation}\label{EvsD}
  \bm E=(\hat{\bm{\epsilon}}^{LL})^{-1} \bm D^L,
\end{equation}
and
\begin{equation}\label{EMvsDM}
  \bm E_{M}=(\hat{\bm{\epsilon}}^{LL}_{M})^{-1} \bm D^L_{M},
\end{equation}
where $\hat{\bm{\epsilon}}^{LL}=\hat {\mathcal P}^L\hat \epsilon\hat {\mathcal
P}^L$ is the longitudinal projection of the dielectric function $\epsilon$
interpreted as a linear operator,
\begin{equation}\label{epsM}
  (\hat{\bm{\epsilon}}^{LL}_{M})^{-1}=\left\langle(\hat{\bm{\epsilon}}^{LL})^{-1}\right\rangle,
\end{equation}
is the inverse of the macroscopic longitudinal dielectric operator,
given\cite{mochanPRB85a, mochanPRB85b} by the spatial average,
$\langle\ldots\rangle$, of the {\em microscopic} inverse longitudinal dielectric
operator, and $\hat{\mathcal P}^L$ is the longitudinal projector operator, which
may be represented in reciprocal space by the matrix
\begin{equation}\label{PL}
  \mathcal P_{\bm G\bm G'}=\hat{\bm  G}\hat{\bm  G}\delta_{\bm G\bm G'},
\end{equation}
with $\bm G$ and $\bm G'$ reciprocal vectors of the metamaterial, where
$\delta_{\bm G\bm G'}$ is Kronecker's delta,
\begin{equation}\label{hatG}
  \hat{\bm G}=\frac{\bm k+\bm G}{||\bm k+\bm G||}
\end{equation}
a unit vector in the direction of the wavevector $\bm k+\bm G$, and $\bm k$ the
conserved Bloch's vector of the linear field which we interpret as the
relatively small wavevector of the macroscopic field.

From Eq. (\ref{epsVsB}) we may write
\begin{equation}\label{epsLL-1}
  (\hat{\bm{\epsilon}}^{LL})^{-1}=\frac{u}{\epsilon_A}(u\hat{\mathcal
    P}^L-\hat B^{LL})^{-1},   
\end{equation}
in which we may interpret the inverse of the operator within parenthesis in
terms of a Green's function,
\begin{equation}\label{Green}
  \hat{\mathcal G}(u)=(u-\hat{\mathcal H})^{-1},
\end{equation}
the resolvent of a Hermitian operator $\hat{\mathcal H}$ with matrix elements
\begin{equation}\label{HGG}
  \mathcal H_{\bm G\bm G'}=\hat{\bm G}\cdot B(\bm G-\bm G')\hat{\bm G}' 
\end{equation}
in reciprocal space, where $B(\bm G-\bm G')$ is the Fourier coefficient of the
periodic characteristic function $B(\bm r)$ with wavevector $(\bm G-\bm G')$.
Notice that $B^{LL}_{\bm G\bm G'}=\hat{\bm G}\mathcal H_{\bm G\bm G'}\hat{\bm
G}'$, $(\bm{\epsilon}^{LL})^{-1}_{\bm G\bm G'}=(u/\epsilon_A)\hat{\bm
G}\hat{\mathcal G}(u)\hat{\bm G'}$, and that
$(\bm{\epsilon}^{LL}_{M})^{-1}=(u/\epsilon_A)\hat{\bm k}\langle\hat{\mathcal
G}(u)\rangle\hat{\bm k}$.

To obtain the macroscopic dielectric response and the microscopic electric field
we proceed as follows. We define a normalized macroscopic state $|0\rangle$ that
represents a longitudinal field propagating with the given small wavevector $\bm
k$ and we act repeatedly on this state with the operator $\hat{\mathcal H}$ to
generate an orthonormal basis set $\{|n\rangle\}$ through
Haydock's\cite{haydock} recursion
\begin{equation}\label{iter}
  \hat{\mathcal H}|n\rangle=b_{n+1}|n+1\rangle+a_{n}|n\rangle+b_{n}|n-1\rangle.
\end{equation}
In this basis, $\hat{\mathcal H}$ may be represented by a tridiagonal matrix
with elements
\begin{equation}\label{Hnn}
  (\mathcal H_{nn'})=\left(
  \begin{array}{ccccc}
    a_0   &b_1   &     0&  0   &\cdots\\
    b_1   &a_1   &b_2   &  0   &\cdots\\
    0     &b_2   &a_2   &b_3   &\cdots\\
    0     &0     &b_3   &a_3   &\cdots\\
    \vdots&\vdots&\vdots&\vdots&\ddots
  \end{array}
  \right)
\end{equation}
given by Haydock's coefficients $a_n$ and $b_n$. Thus, the macroscopic inverse
longitudinal response may be obtained as a continued fraction
\cite{mochanOE10,perezNJP13}
\begin{equation}\label{epsMH}
\begin{split}
  (\hat{\bm{\epsilon}}^{LL}_{M})^{-1}
&= \hat{\bm k}\hat{\bm k}\frac{u}{\epsilon_A}\langle0|(u-\hat H)^{-1}|0\rangle\\
&= \hat{\bm k}\hat{\bm k}\frac{u}{\epsilon_A}
   \frac{1}{u-a_0-
      \frac{b_1^2}{u-a_1-
        \frac{b_2^2}{u-a_2-
          \frac{b_3^2}{\ddots}}}}      
\end{split}
\end{equation}
and the microscopic electric field (\ref{EvsD}) may be represented in reciprocal
space by
\begin{equation}\label{field}
  E_G=\sum \zeta_n \langle\bm G|n\rangle
\end{equation}
with coefficients $\zeta_n$ obtained by solving the tridiagonal system
\begin{equation}\label{zeta_n}
  \sum_{n'} (u\delta_{nn'}-H_{nn'}) \zeta_{n'}=\delta_{n0} D^L,
\end{equation}
where we write the fields in real space as
\begin{equation}\label{Dvsk}
  \bm D^L(\bm r)=\hat{\bm k} D^L e^{i\bm k\cdot\bm r}
\end{equation}
and
\begin{equation}\label{EvsG}
  \bm E(\bm r)=\sum_{\bm G} \hat{\bm G} E_G e^{i(\bm k +\bm G)\cdot\bm r}.
\end{equation}

Notice that the results of the calculation above depend on the direction
$\hat{\bm k}$ chosen as the propagation direction of the external field. As we
may identify
\begin{equation}\label{epsvsk}
  (\bm{\hat \epsilon}_{M}^{LL})^{-1}=\frac{\hat{\bm k}\hat{\bm
      k}}{\hat{\bm k}.\bm{\hat \epsilon}_{M}^{LL}\cdot\hat{\bm k}},
\end{equation}
all the components of the macroscopic dielectric tensor may be efficiently
obtained from Eq. (\ref{epsMH}) by repeating the calculation of its longitudinal
proyection for different propagation directions $\hat{\bm k}$, such as along all
independent combinations $\hat{\bm e}_i+\hat{\bm e}_j$ of pairs of cartesian
directions $\hat{\bm e}_i$ and $\hat{\bm e}_j$ ($i,j=x,$ $y$ or $z$).

Once we obtain the microscopic field from Eqs. (\ref{field}), (\ref{zeta_n}) and
(\ref{EvsG}), we may substitute it in Eqs. (\ref{Pf})-(\ref{Q}) to obtain the
forced SH polarization, which we may then substitute in Eq. (\ref{P}) to obtain
the self-consistent quadratic polarization in the SH. However, in order to solve
Eq. (\ref{P}) we need the self-consistent SH field, which in the long wavelength
approximation is simply given by the depolarization field
\begin{equation}\label{depol}
  \bm E({2\omega})=-4\pi\bm P^L({2\omega})
\end{equation}
produced only by the longitudinal part of the SH polarization. Thus we write Eq.
(\ref{P}) as
\begin{equation}\label{PvsPL}
  \bm P({2\omega})=-4\pi n\alpha(2\omega) \bm P^L({2\omega}) +\bm
  P^f({2\omega}). 
\end{equation}
By taking its longitudinal projection,  we obtain a
closed equation for $\bm P^L({2\omega})$ which we solve formally as
\begin{equation}\label{P2L}
  \bm P^L({2\omega})=(\hat{\bm\epsilon}^{LL}(2\omega))^{-1}\bm
  P^{fL}({2\omega})   
\end{equation}
using Eq. (\ref{eps}). Plugging this result back into Eq. (\ref{PvsPL}), we
finally obtain the SH polarization $\bm P(2\omega)$.

In order to perform the operation indicated in Eq. (\ref{P2L}) we perform a
Haydock recursion as in Eq. (\ref{iter}) but using $\bm
P^{fL}{(2\omega)}$ to construct a new initial normalized state
$|\tilde 0\rangle$, with components $\langle \bm G|\tilde 0\rangle$ in
reciprocal given by
\begin{equation}\label{PfvsG}
  \bm P^{fL}_{\bm G}({2\omega})=\hat{\bm G} \langle
  \bm G|\tilde 0\rangle f,
\end{equation}
where $f$ is a normalization constant. From this state, we build a new Haydock
orthonormal 
basis $|\tilde n\rangle$ using the same procedure as in Eq. (\ref{iter}). Thus,
we write the self-consistent longitudinal SH polarization as
\begin{equation}\label{Plvsr}
  \bm P^L({2\omega};\bm r)=\sum_{\bm G} P^L_{\bm G}(2\omega) \hat{\bm G} 
  e^{i(\bm k +\bm G)\cdot\bm r}.   
\end{equation}
with
\begin{equation}\label{PLvsn}
  P^L_{\bm G}{(2\omega)}=\frac{u_2}{\epsilon_{A2}}\sum_{\tilde n} \xi_{\tilde n} \langle\bm G|\tilde n\rangle
\end{equation}
and with coefficients $\xi_{\tilde n}$ obtained by solving the tridiagonal system
\begin{equation}\label{xin}
  \sum_{\tilde n'} (u_{2}\delta_{\tilde n\tilde n'}-H_{\tilde n\tilde
    n'}) \xi_{\tilde n'}=\delta_{\tilde n\tilde 0} f,
\end{equation}
where $u_{2}$ and $\epsilon_{A2}$ are the spectral variable (\ref{u}) and the
dielectric response $\epsilon_A$  but evaluated at the SH frequency $2\omega$.

Substitution of $\xi_{\tilde n}$ from Eq. (\ref{xin}) into Eqs. (\ref{PLvsn})
and (\ref{Plvsr}) yields the SH longitudinal polarization, which may then be
substituted into Eq. (\ref{PvsPL}) to obtain the total SH polarization in the
longwavelength limit when the system is excited by a longitudinal external field
along $\hat{\bm k}$. Averaging the result, or equivalently, taking the $\bm G=0$
contribution in reciprocal space, we obtain the macroscopic SH polarization $\bm
P_{M}(2\omega)$ which we write as
\begin{equation}\label{P2M}
  \bm P_{M}({2\omega})=\frac{1}{4\pi}(\bm\epsilon_{M}(2\omega)-\bm 1)\bm
    E_{M}({2\omega})+\bm P_{M}^f({2\omega}),
\end{equation}
where the first term is the contribution of the linear response at
$2\omega$ to the SH
macroscopic field, and the second term
\begin{equation}\label{P2Mf}
  \bm P_{M}^f({2\omega})=\bm \chi^{(2)}_{M}:\bm E_{M}({\omega})\bm E_{M}({\omega})  
\end{equation}
is the sought after contribution to the SH macroscopic polarization
forced by the 
fundamental macroscopic 
electric field, and $\bm \chi^{(2)}_{M}$ is the corresponding SH quadratic
macroscopic susceptibility, given by a third rank tensor. Within our
longwavelength longitudinal calculation the macroscopic field $\bm
E_{M}({2\omega})$ is simply given by the longitudinal depolarization field
\begin{equation}\label{E2MvsP2LM}
  \bm E_{M}({2\omega})=\bm E^L_{M}({2\omega})=-4\pi \bm P^L_{M}({2\omega}),
\end{equation}
so that, taking the longitudinal projection of Eq. (\ref{P2M}) we
obtain
\begin{equation}\label{P2ML}
  \bm P_{M}^{fL}({2\omega})=\hat{\bm k}\hat{\bm k}\cdot \bm P_{M}^f({2\omega}) =\bm\epsilon_{M}^{LL}(2\omega) \bm P_{M}^{L}({2\omega}).
\end{equation}
Substituting $\bm P_{M}^{fL}({2\omega})$ from Eq. (\ref{P2ML}) into
(\ref{E2MvsP2LM}) and then into (\ref{P2M}) we obtain the macroscopic forced
quadratic SH polarization $\bm P_{M}^f(2\omega)$ produced by a longitudinal
external $\bm D^L$ field pointing along $\hat{\bm k}$. As in the linear case, we
finally repeat the calculation above, for several independent directions of
propagation $\hat{\bm k}$ so that the corresponding Eqs. (\ref{P2Mf}) become a
system of linear equations in the unknown cartesian components
$\chi^{(2)}_{M\,ijk}$ ($i,j,k=x,$ $y,$ or $z$) which we solve to obtain the
third rank second order susceptibility tensor $\bm\chi^{(2)}_{M}$ of the
metamaterial.

In summary, to obtain the quadratic response we first obtain the nonretarded
microscopic field and the macroscopic dielectric tensor using a Haydock's
recursion starting from a macroscopic external longitudinal field, then we use
the dipolium model to obtain the microscopic {\em source} of the SH
polarization, 
we screen it using Haydock's scheme again to obtain the {\em full} microscopic
polarization, which we average to obtain the full macroscopic SH polarization.
As this {\em includes} a contribution from the {\em macroscopic SH
depolarization field}, we 
substract it before identifying the quadratic suceptibility tensor projected
onto the longitudinal direction. We repeat the calculation along different
independent directions so that we can extract all the components of the
quadratic susceptibility.

In the process above we assumed that the unit cell of the metamaterial is small
with respect to the wavelength at frequency $\omega$, and thus we introduced a
long-wavelength approximation and assumed the external field and the electric
field to be longitudinal. After obtaining all the components of the macroscopic
response, we should not concern ourselves anymore with the texture of the
metamaterial; the unit cell disappears from any further use we give to the
macroscopic susceptibility. Thus, we can solve any macroscopic SH related
electromagnetic problem using the suceptibility obtained above without using
again the long wavelength approximation. Once we have the full
macroscopic susceptibility tensor we may use it to calculate the
response to transverse as well as longitudinal fields. 
Thus, we may use our susceptibility
above to study the generation of electromagnetic waves at the SH from a
propagating fundamental wave, in which case the macroscopic fields
{\em can  no
longer} be assumed to be longitudinal.

%%%%%%%%%%%%%%%%%%%%%%%%%%%%%%%%%%%%%%%%%%%%%%%%%%%%%%%%%%%%%%%%%%%%%%%%%%%%%%%%
%%%%%%%%%%%%%%%%%%%%%%%%%%%%%%%%%%%%%%%%%%%%%%%%%%%%%%%%%%%%%%%%%%%%%%%%%%%%%%%%

\section{Results}\label{sec:results}

We present results for a simple geometry in which we can control the degree of
centrosymmetry. To that end, we  incorporated the scheme described in the
previous section into the package {\em Photonic} \cite{photonic}, which is a
modular, object oriented system based on the Perl programming language, its
Perl Data Language (PDL) \cite{glazebrook97pdl} extension for efficient
numerical calculations, and the Moose \cite{moose} object system. The package
implements Haydock's recursive procedure to calculate optical properties of
structured metamaterials in the nonretarded as well as in the
retarded regime.

\begin{figure}
\includegraphics[width=0.6\linewidth]{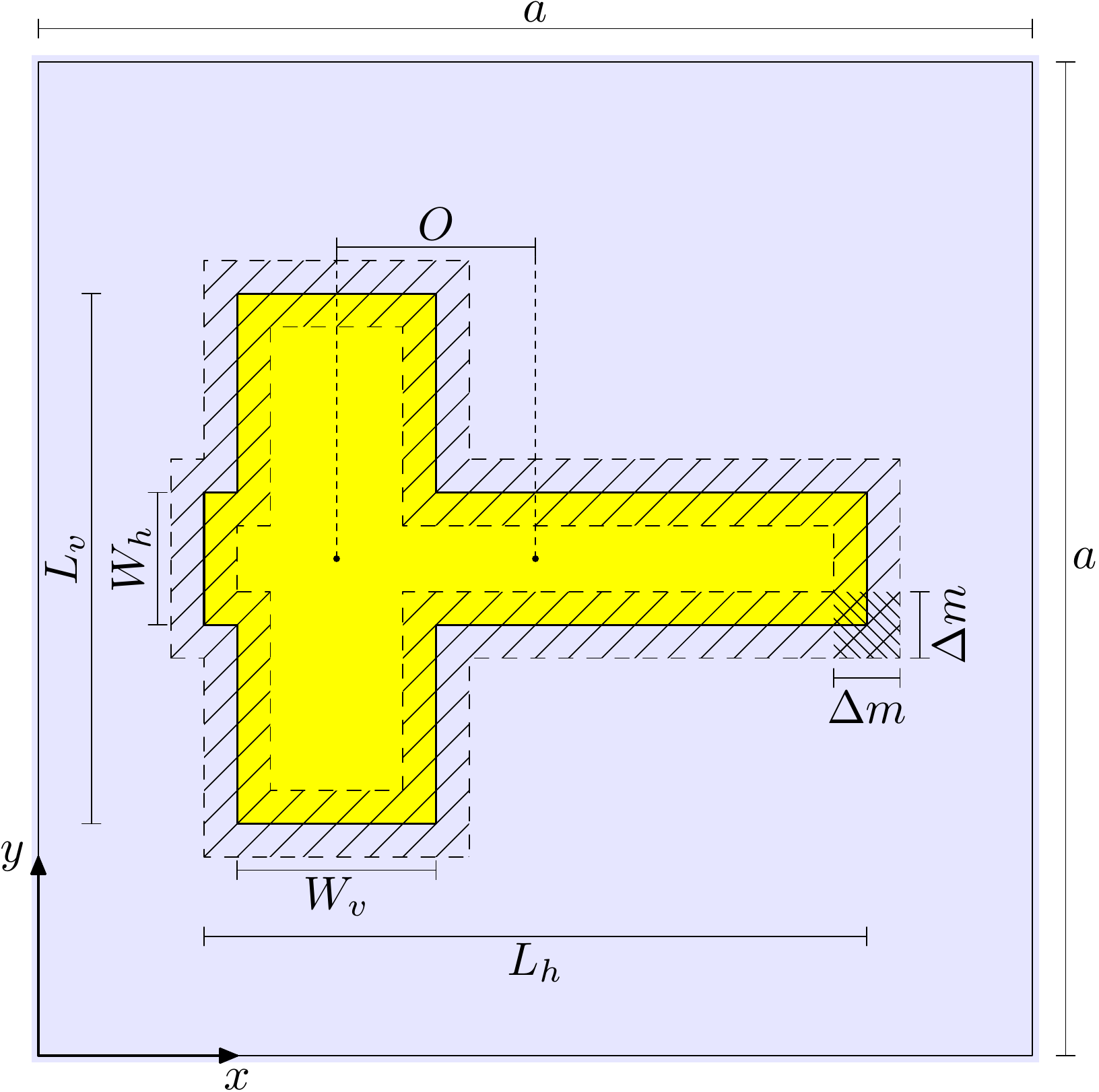}
  \caption{\label{fig-1}Unit cell of a metamaterial made up of a
    horizontal and a 
    vertical rectangular hole within a conducting matrix. We indicate
    the lattice parameter $a$ of the square array, the length
    $L_\beta$ and width $W_\beta$ of each rectangle 
    ($\beta=h, v$) and the offset $O$ of the center of the vertical
    rectangle with 
    respect to that of the horizontal one. We indicate the
    directions $x$, $y$ of the crystalline axes. The shaded regions
    correspond to masks of width $\Delta m$ used to single out the
    surface, edge and corner contributions to the SH response. }
\end{figure}

Our system consists of a square array of pairs of holes in the
shape of prisms with a rectangular cross section within a metallic host (Fig.
\ref{fig-1}).
Each rectangle is aligned with one of the crystalline axes $x$, $y$ of the
metamaterial and is characterized by its length $L_h$ or $L_v$ and its width
$W_h$ or $W_v$, where $h$ denotes horizontal (along $x$) and $v$ vertical (along
$y$) alignment. The center of the vertical rectangle is shifted horizontally
with respect to the center of the horizontal rectangle by an offset $O$. Thus,
when $O=0$ our system is centrosymmetric and as $O$ increases it becomes
noncentrosymmetric in varying degrees.

\begin{figure}
\centering
\includegraphics[width=0.7\linewidth]{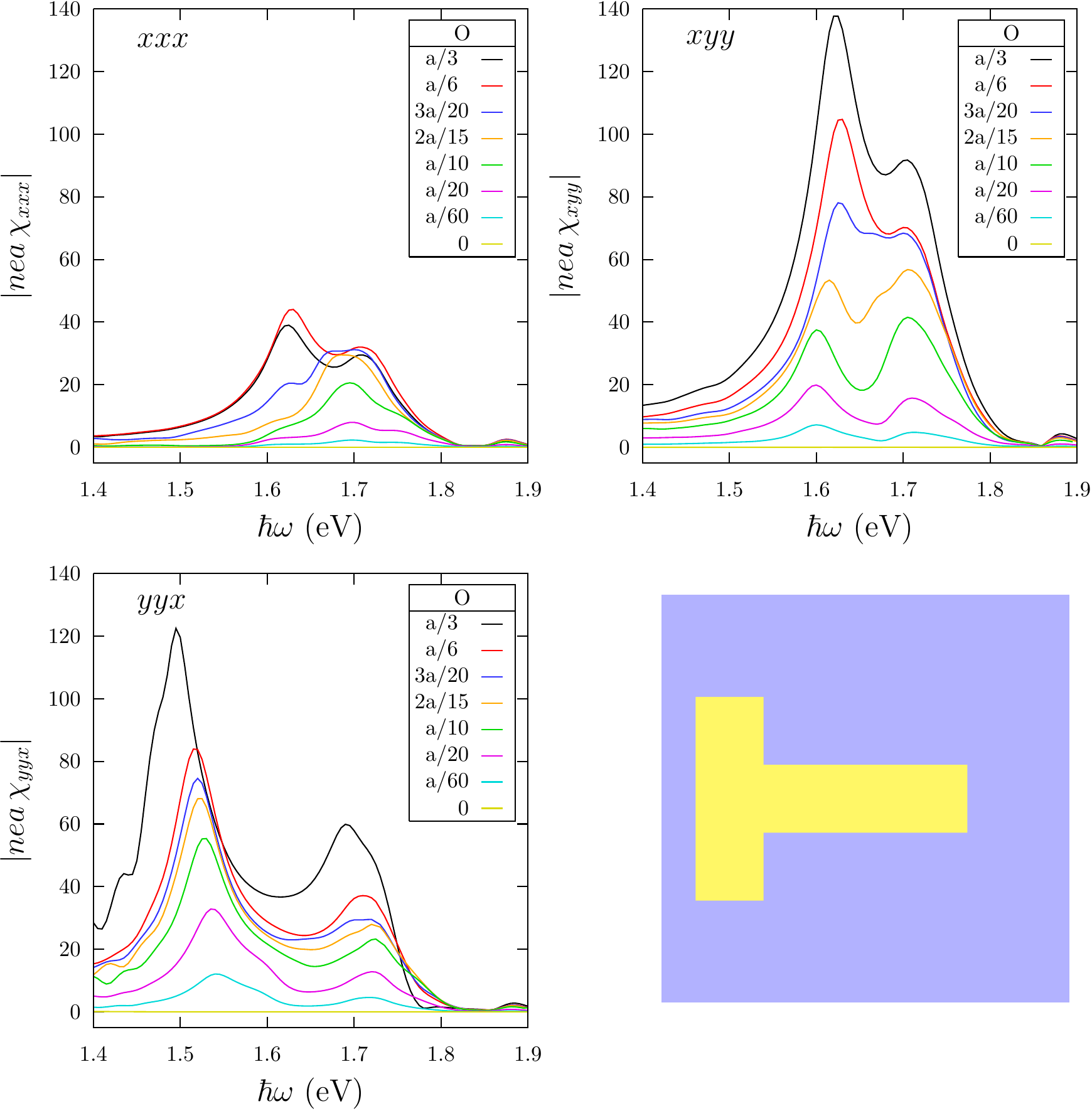}
  \caption{\label{fig-2} Normalized absolute value of the non-null
    components of the SH susceptiblity $nea\chi_{ijk}$, with
    $ijk=xxx$ (upper left), $xyy$ (upper right), and $yyx=yxy$ (lower
    left), for a square
    lattice of rectangular holes, as in Fig. \ref{fig-1} within an Ag
    matrix,
    with geometrical parameters $L_h=L_v=a/2$, $W_h=W_v=a/6$,
    for different values of the offset
    $O=0\ldots a/3$. The lower right panel displays the geometry
    corresponding to the largest offset. Notice that for these cases
    the holes overlap.
 }
\end{figure}

In order to simplify our analysis, we have chosen a system that has mirror
symmetry $y\leftrightarrow -y$.
Thus, the only in-plane non-null components of the SH susceptibility
are\cite{popovbook} $\chi_{xxx}$, $\chi_{xyy}$, and $\chi_{yxy}=\chi_{yyx}$. We
omit the subindex $M$ and the superindex $(2)$ that indicate these are
components of the quadratic macroscopic susceptibility in order to simplify the
notation, as we expect it yields no confusion.
In Fig. \ref{fig-2} we show the spectra
of the magnitude of these non-null components for an Ag host\cite{yangPRB15} and
for different values of the offset $O$. The parameters we used were
$W_h=W_v=a/6$, $L_h=L_v=a/2$, $O=0\ldots a/3$. 
Notice that when $O=0$ the system is centrosymmetric and there is no SH signal.
As $O$ increases towards $\pm a/3$ the system becomes noncentrosymmetric. Two
resonances become clearly visible and they grow in size as $O$ increases and the
system moves farther away from the centrosymmetric case. The lower energy
resonance of $\chi_{yyx}$ is at a different frequency than those of $\chi_{xxx}$
and $\chi_{xyy}$ and is red shifted as the offset increases. If $O$ increases
beyond $a/3$ (not shown) the two rectangles would cease to overlap and the
quadratic suceptibility would rapidly decay, until $O=a/2$ for which the system
becomes exactly centrosymmetric again and the quadratic susceptibility becomes
exactly null.

According to Fig. \ref{fig-2}, the order of magnitude of the SH
susceptibility is around 
$10^2/ nea$. For typical noncentrosymmetrical materials, such as quartz, the
corresponding order of magnitude is about $1/nea_B$, where $a_B$ is Bohr
radius\cite{boydbook}. Thus, a centrosymmetric material with a
noncentrosymmetric geometry can achieve susceptibilities of the order of $10^2
a_B/a$ times that of noncentrosymmetrical materials. Thus, quadratic
metamateriales made of centrosymmetrical materials may be competitive as long as
the lattice parameter is not too large.

\begin{figure}
\centering
\includegraphics[width=0.5\linewidth]{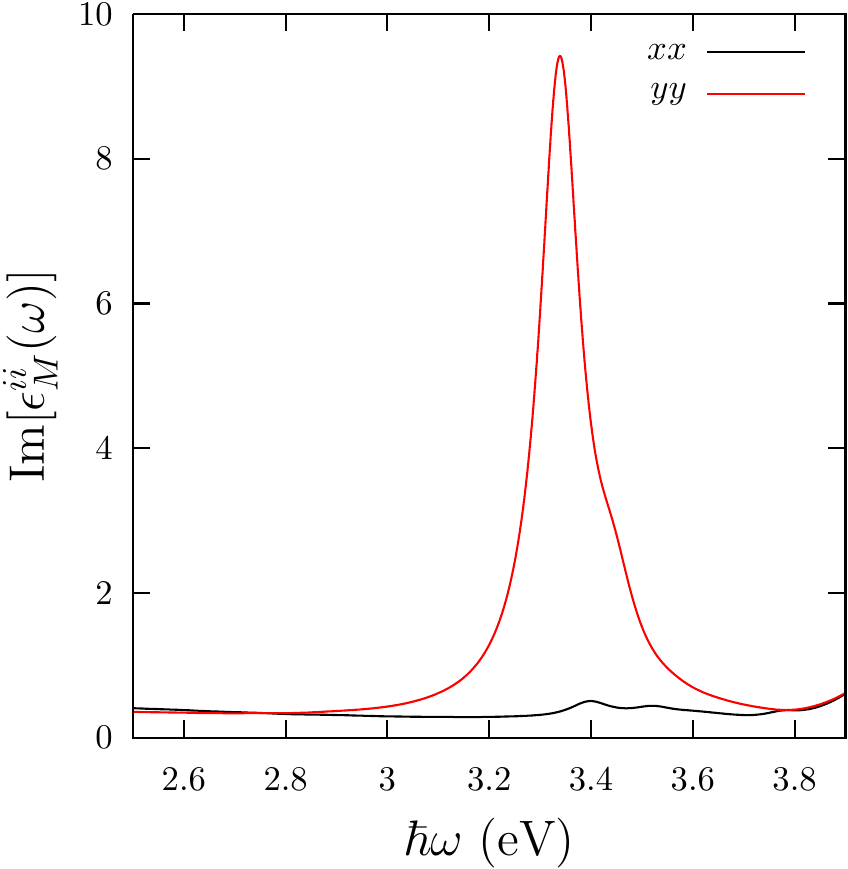}
  \caption{\label{fig-3} Non-null components of the macroscopic
    dielectric response, $\epsilon_{M}^{xx}$ and $\epsilon_{M}^{yy}$, of a 
    metamaterial made up of a square array of horizontally oriented
    single rectangular holes with the same dimensions as in
    Fig. \ref{fig-2} within an Ag matrix.}
\end{figure}

\begin{figure}
\centering
\includegraphics[width=0.9\linewidth]{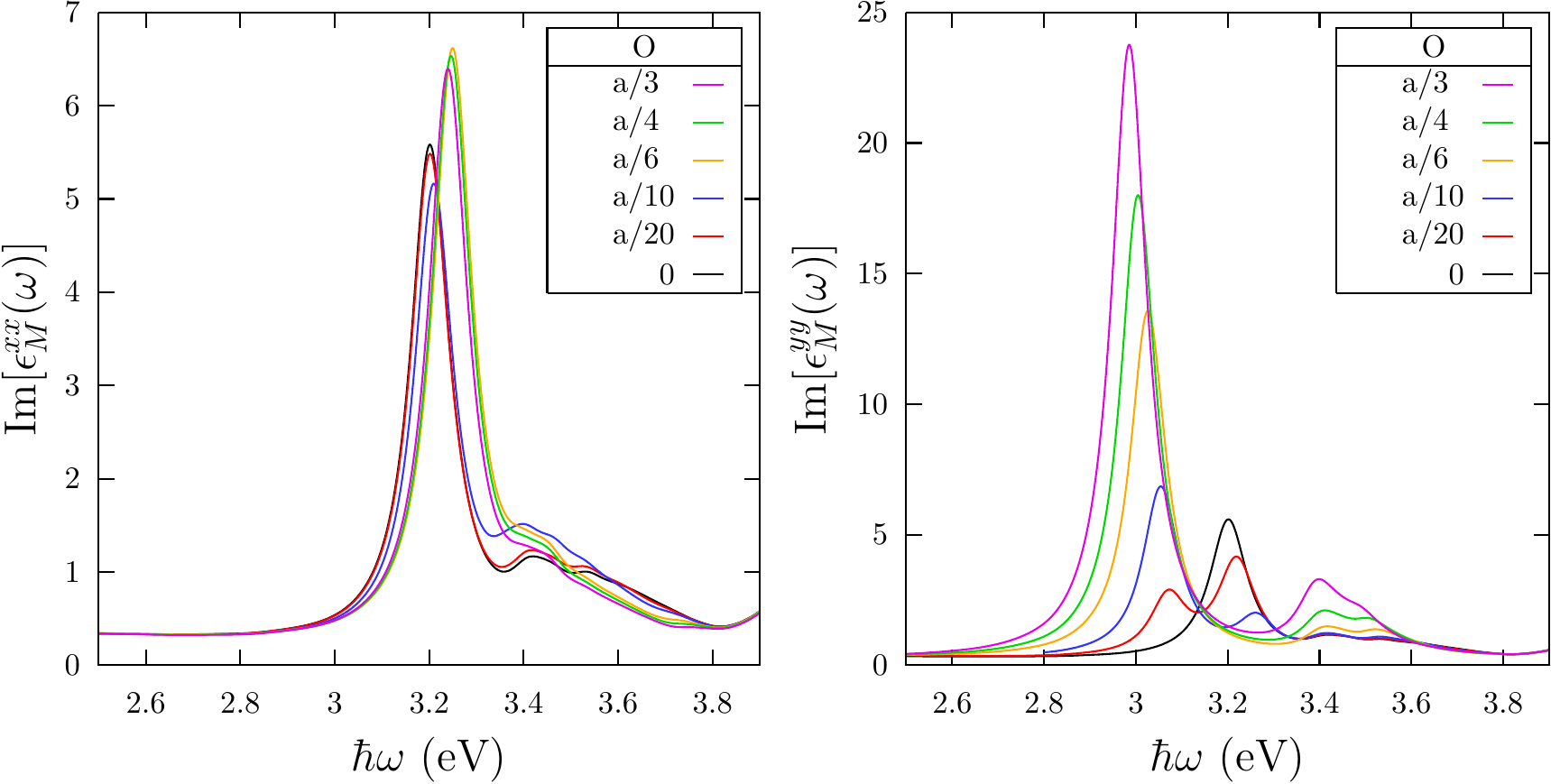}  
\caption{\label{fig-4} Non-null components $\epsilon_{M}^{xx}$ and
    $\epsilon_{M}^{yy}$ of the macroscopic
    dielectric tensor $\bm\epsilon_{M}$ of a
    metamaterial made up of a square lattice of pairs of horizontally
    and vertically oriented
    single rectangular holes within an Ag matrix as in
    Fig. \ref{fig-1} with the same parameters as in Fig. \ref{fig-2}
    for different 
    values of the offset $O=0\ldots a/3$.}
\end{figure}

In order to understand the origin of the structure of the spectra discussed
above, in Fig. \ref{fig-3} we plot the non-null components $\epsilon_{M}^{xx}$
and $\epsilon_{M}^{yy}$ of the macroscopic linear dielectric tensor
$\bm\epsilon_{M}$ of a metamaterial made up of a square lattice of
{\em single}
rectangular holes with a horizontal orientation.
Notice that there is a very weak resonance
close to 3.4\,eV  corresponding to polarization along the length of
the rectangle 
($x$ direction) and a strong resonance corresponding to polarization along the
width of the rectangle ($y$ direction) at a slightly smaller frequency.
Although there is a strong linear resonance in the $y$ direction, this system is
centrosymmetrical and would yield no SH signal. When we combine horizontal and
vertical rectangles (Fig. \ref{fig-4}) with a null offset $O=0$ to
make a centrosymmetric array of 
crosses, both resonances appear for both polarizations, although they now
interact, partially exchange their strengths and repel so that both become
clearly visible close to 3.4\,eV and 3.2\,eV.

\begin{figure}
\centering
\includegraphics[width=0.7\linewidth]{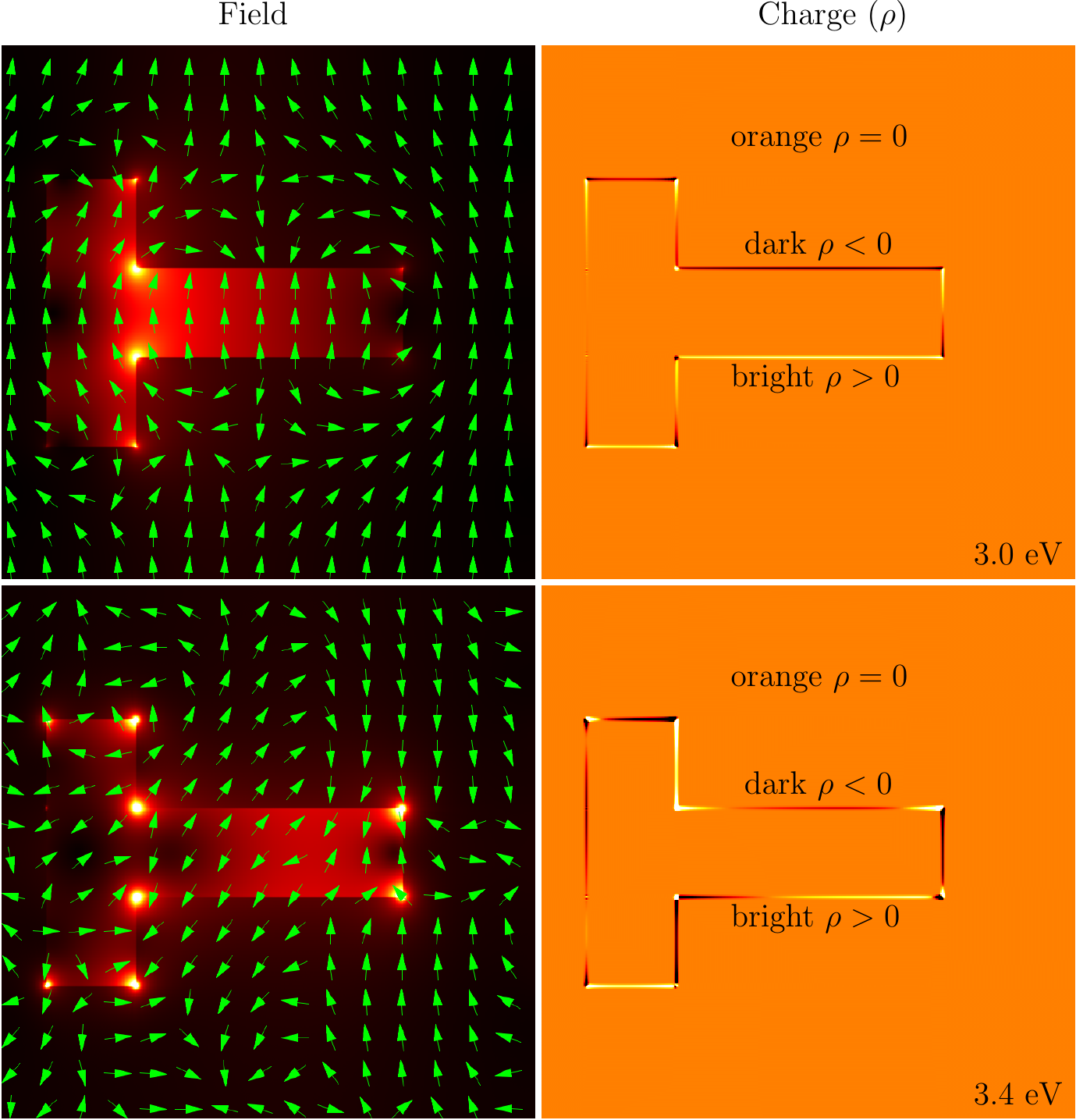}  
\caption{\label{fig-5}Magnitude (color coded) and direction
  (arrows) of the microscopic linear electric field  (left) and
    induced charge density $\rho$ (right) for a metamaterial made of a square
    lattice of rectangular holes within an Ag
    matrix with the same parameters as in Fig. \ref{fig-2} with an
    offset $O=a/3$, excited by a
    macroscopic field along the $y$ (vertical) direction 
    for $\hbar\omega\approx 3\,eV$ and $\hbar\omega\approx 3.4\,eV$ 
    corresponding to the two peaks
    in $\epsilon_{M}^{yy}$ shown in Fig. \ref{fig-4}. The field
    and the charge distribution correspond to a vertical polarization
    for the horizontal rectangle, a vertical polarization for the
    vertical rectangle and a nondiagonal quadrupole with opposite
    horizontal polarizations above and below the symmetry plane. 
  }
\end{figure}

As the offset $O$ increases, there are only small changes to
the spectra corresponding to $\epsilon_{M}^{xx}$, consisting in
changes to the weights of the peaks. However, a new strong mode develops
in the spectra of   $\epsilon_{M}^{yy}$. This mode is due to the strong coupling
of a quadrupolar oscillation in the vertical rectangle to the vertical dipolar
oscillation of the horizontal rectangle. This quadrupole may be visualized as a
horizontal polarization in the upper part of the vertical rectangle and a
horizontal polarization in the opposite direction in the lower part of the
rectangle, as illustrated by Fig. \ref{fig-5}. The coupling is symmetry allowed
as for a finite offset $O\ne0$ the system looses the $x\leftrightarrow -x$
symmetry.

We expect the resonant structure of the quadratic susceptibility to have peaks
corresponding to the resonances of the linear response at the
fundamental and at 
the SH frequency. Thus, we expect peaks at the fundamental and at the
subharmonics of those of the linear response. As there is no
structure in the linear response within the region from 1.4\,eV to
1.9\,eV shown in Fig. 
\ref{fig-2}, in our system we can only expect structure at the
subharmonics, due 
to a resonant excitation of the polarization at the SH frequency. For a
macroscopic field oriented along the cartesian directions $x$ or $y$ the SH
harmonic polarization can only point along the $x$ direction, due to the
$y\leftrightarrow -y$ mirror symmetry of our system. Thus, the subharmonics of
the resonances of $\epsilon_{M}^{xx}$ (Fig. \ref{fig-4}) appear in the
susceptibility components $\chi_{xxx}$ and $\chi_{xyy}$ (Fig. \ref{fig-2}). On
the other hand, a macroscopic field that points along an intermediate direction
between $x$ and $y$ may excite a quadratic polarization along $y$. Thus, the
subharmonics of the resonances of $\epsilon_{M}^{yy}$ (Fig. \ref{fig-4}) appear
in the susceptibility components $\chi^{yxy}=\chi^{yyx}$ (Fig. \ref{fig-2}).

\begin{figure}
\includegraphics[width=0.85\linewidth]{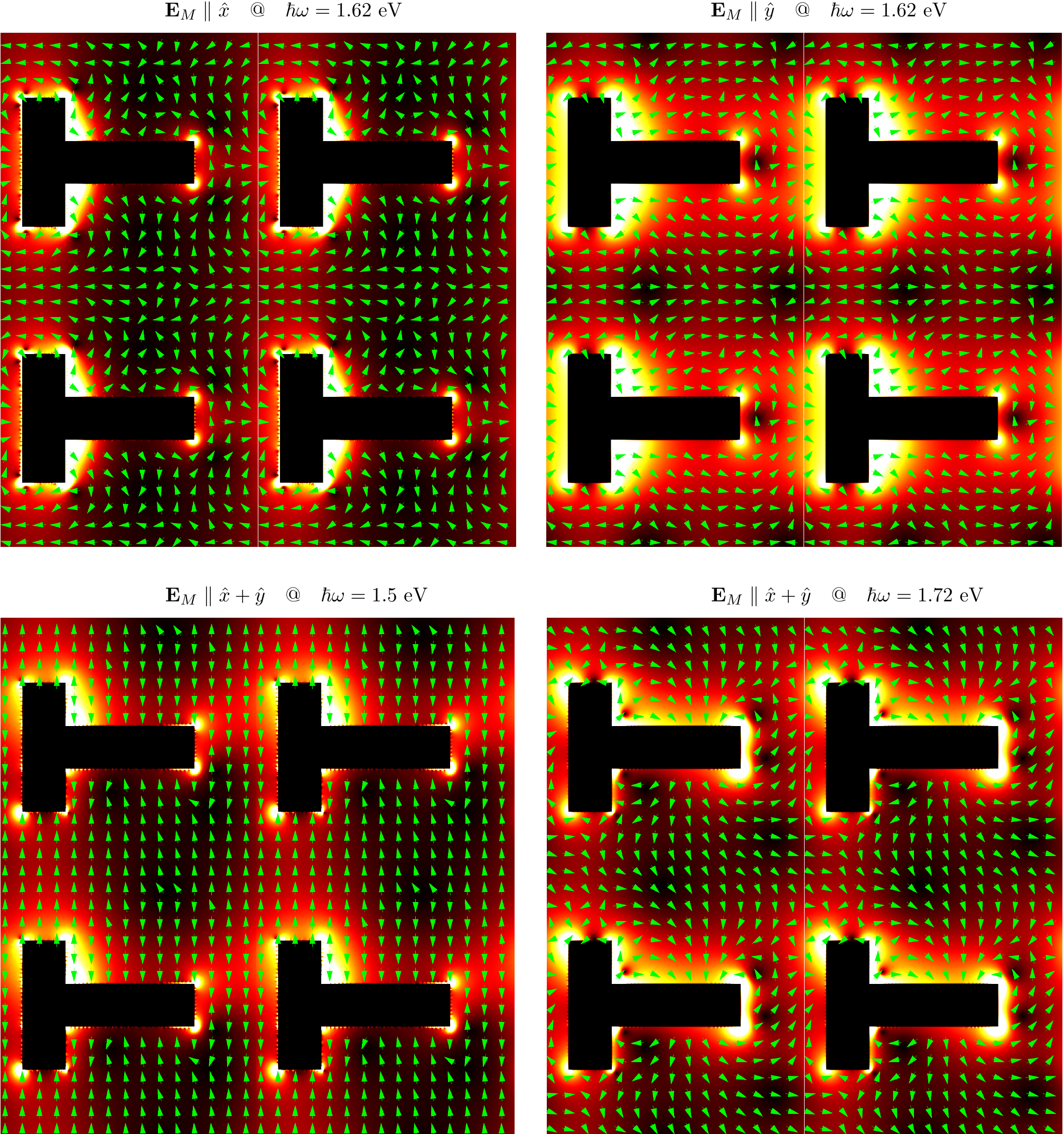}
  \caption{\label{fig-6}Magnitude and direction of the quadratic
    polarization induced in the same system as in Fig. \ref{fig-2}
    for the largest 
    offset $O=a/3$ at the resonant energies
    $\hbar\omega=1.62\,eV$ and the fundamental macroscopic field $\bm
    E_{M}$ along the direccion $\hat x$ (upper left), 
    $\hbar\omega=1.62\,eV$ and $\bm E_{M}$ along $y$ (upper right) and
    for $\hbar\omega=1.5\,eV$ and  $\hbar\omega=1.72\,eV$ with
    $\bm E_{M}$ along $\hat x+\hat y$ (bottom).
}
\end{figure}

To gain further insight into the nature of the resonances, in Fig. \ref{fig-6}
we show the polarization maps evaluated at the maxima of the SH spectra
corresponding to different directions of the macroscopic linear field, and for
the offset $O=a/3$ that yields the largest signals.
We notice that when the fundamental macroscopic field points along the
$x$ or along the $y$
direction, the magnitude of the SH polarization is symmetric with respect to the
mirror plane, the $y$ component of the polarization points towards
opposite directions on either side of the mirror plane, yielding a macroscopic
SH polarization along $x$. In these cases, the polarization has maxima near the
four concave vertices of the vertical hole and near the convex 
vertex where the horizontal and vertical rectangles meet. On the other
hand, when the fundamental 
macroscopic field points along the direction of $\hat x+\hat y$, the resulting
quadratic polarization has no symmetry at all, and it yields a macroscopic SH
polarization that has a $y$ component.

\begin{figure}
\includegraphics[width=0.5\linewidth]{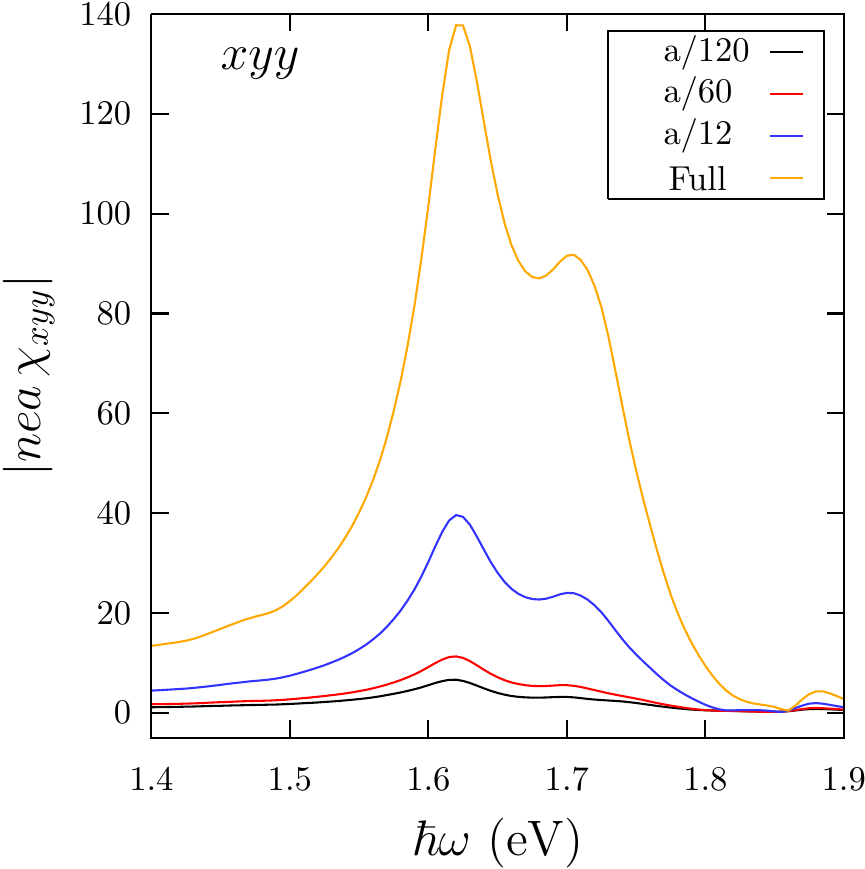}
\caption{\label{fig-7}Contributions to the quadratic susceptibility
    $\chi_{xyy}$ of the same system as in Fig. \ref{fig-6} from the
    region within a distance $\Delta m$ from the surface, as defined in
    Fig. \ref{fig-1} for various values of $\Delta m=a/120, a/60,
    a/12$, and the full susceptibility. 
}
\end{figure}

Finally, in Fig. \ref{fig-7} we illustrate the contributions of the surface
region to the total quadratic susceptibility by adding only the contributions
within bands of varying widths $\Delta m$ around the surface.
We notice that although there is a very strong surface polarization, its
contribution to the macroscopic quadratic susceptibility is relatively 
small, as it is confined to a very narrow region and it is partially
cancelled by the polarization at other parts of the surface, so that
for the geometry studied here, most of the SH signal comes from the
bulk of the host.

\section{Conclusions}\label{sec:conc}

We have developed a formalism for the calculation of the second order
susceptibility of structured binary metamaterials formed by a lattice of
particles embedded within a host, for the case where both components
consists of centrosymmetric materials but where the geometry is not
centrosymmetric. Although SH is strongly suppressed within a
homogeneous centrosymmetric material, the noncentrosymmetric surface
is capable of sustaining a surface nonlinear polarization and to
induce a strongly varying linear field which induces a multipolar
nonlinear polarization within the metamaterial components.

We implemented our formalism using the Haydock recursive scheme
within the {\em Photonic} modular package and
applied it to the calculation of
the second-order nonlinear susceptibility of a structured
metamaterial composed of a homogeneous Ag host with a
lattice of pairs of rectangular holes. By
modifying the geometry of the holes, we modify the degree of
non-centrosymmetry of the material, allowing us to fine-tune both the
peak position and intensity of the SH response. The SH
signal is very sensitive 
to changes in the geometrical parameters of the structure.

After establishing the inclusion shape that most enhances this signal,
we analyzed the polarization field and showed that the SH response is
largest at resonance close to the concave and convex corners but it
extends well into the host material. The order of magnitude of the
susceptibility obtained in this calculation is comparable to that of
typical non-centrosymmetric materials.

Although this study was carried out for one particular combination of materials,
the employed procedure is equally valid for calculating the nonlinear properties
for any metamaterial composed of arbitrary materials and inclusions. Only
\emph{a priori} knowledge of the dielectric function of each constituent
material is required. This approach affords the opportunity to quickly and
efficiently study a limitless range of possible metamaterial designs,
with manifold optical applications in mind. Our hope is that this
methodology will prove to be an important tool for future metamaterial design
and fabrication.

%%%%%%%%%%%%%%%%%%%%%%%%%%%%%%%%%%%%%%%%%%%%%%%%%%%%%%%%%%%%%%%%%%%%%%%%%%%%%%%%
%%%%%%%%%%%%%%%%%%%%%%%%%%%%%%%%%%%%%%%%%%%%%%%%%%%%%%%%%%%%%%%%%%%%%%%%%%%%%%%%

\acknowledgments

This work was supported by DGAPA-UNAM under grants IN113016 and
IN111119 (WLM) and 
by CONACyT  under scholarship 589138 (URM). We acknowledge useful
talks with Raksha Singla and Sean M. Anderson.

\bibliography{cites.bib}

\end{document}